# AN FFAG ACCELERATOR FOR THE INTENSITY FRONTIER*

L. Jenner, C. Johnstone, D. Neuffer, Fermilab, Batavia 60510, USA
J. Pasternak, Imperial College, London,/STFC-RAL, ISIS, Harwell, UK

*Abstract*

The next generation of high-energy physics experiments requires high intensity protons in the multi-GeV energy range for efficient production of secondary beams. The Fermilab long-term future requires an 8 GeV proton source to feed the Main Injector for a 2 MW neutrino beam source in the immediate future and to provide 4 MW pulsed proton beam for a future neutrino factory or muon collider. We note that a 3GeV cw linac matched to a 3---8GeV FFAG ring could provide beam for both of these mission needs, as well as the cw 3GeV experiments, and would be a natural and affordable scenario. We present details of possible scenarios and outline future design and research directions.

## INTRODUCTION

The presently developing scenario for Project X at Fermilab is based on a ~3GeV cw linac, providing beam at 1mA [1]. This linac must also feed an accelerator that takes beam to ~8GeV for injection into the Main Injector (MI), and that accelerator must be upgradeable to provide bunched ~8GeV beam at 4MW for future needs. A rapid-cycling synchrotron or pulsed linac is being considered for the Main Injector; but a synchrotron would not be upgradeable and a pulsed linac is mismatched to the cw source.

An FFAG (fixed-field alternating gradient) accelerator could accelerate beam from 3 to 8GeV using a few MV of ramped RF much like a synchrotron but without ramped magnets; the magnets have an enlarged size needed to accommodate stable orbits over the full momentum range. However upgrade to the requirements of the pulsed facility (factory or collider) is relatively straightforward. With fixed-field magnets, the RF can be changed to higher-power faster-pulsing capacity, with RF frequencies changed to fit the different bunching requirements. The enlarged apertures needed to accommodate the 3 to 8 GeV momentum change may also accommodate the larger transverse emittances needed to obtain the pulsed bunches.

## FFAG-BASED 3-8 GEV SCENARIO

FFAG design has progressed dramatically over the last few years. Y. Mori initiated this renaissance by building and operating a number of scaling FFAG designs [2]. In a radially scaling FFAG, the fields strictly increase with radius, following $B(r, z) = B_0(z) (r/r_0)^k$, where k is the FFAG field index. This forces the beam dynamics to scale with momentum, and naturally obtains constant tune with momentum. More recently. "non-scaling" FFAG designs have been developed. Initially these designs had a large tune dependence with momentum, which limited beam lifetimes to a few turns [3]. Following this, non-scaling designs with constant or small-variation in tunes are being developed, and these can have the many-turn stability needed for a proton driver. [4, 5] These non-scaling solutions depart from strict adherence to the scaling law but approximate scaling beam dynamics.

*Beam requirements*

In the near-term future, the MI will be used to provide 2 MW beam at 60 GeV (~1.25Hz pulses) or 120 GeV (~0.67Hz) for long baseline neutrino beams, and thus will need to be filled with ~1.6 × $10^{14}$ protons/pulse. The injection energy would be 8 GeV, matching the present cycle, and enabling the use of the Recycler as an intermediate stacking ring. This would require ~26ms of 1mA beam from the cw linac, with particle energies boosted to 8 GeV by the FFAG. 6 cycles of an ~500m circumference FFAG (3→8GeV) cycling at 10Hz or more, with each cycle accelerating ~2.7×$10^{13}$ p, could fill the 3320m circumference Recycler for transfer to the MI. 4½ ms of charge exchange injection from the cw linac would be needed for the FFAG in each cycle.

In the far-term future, an ~8 GeV proton source must provide 4 MW intense beam with pulses of beam at ~60 to ~15 Hz for a neutrino factory (NF) or muon collider (MC). A 3→8 GeV FFAG system producing 4 bunches of ~1.3 × $10^{13}$ p at 60Hz (NF mode) or 4 bunches of 5 × $10^{13}$ p at 15 Hz (MC mode) would meet these requirements. For these modes, 50% of the 3GeV 1mA cw beam would be needed for injection; a 3GeV accumulator ring accepting 8.3 mA-ms (NF) or 33mA-ms (MC) of cw H$^-$ linac injected beam could be used. Protons would be transferred from the accumulator to the FFAG for 3 to 8 GeV acceleration. The protons would need to be formed into 4 short bunches (~3ns); this formation could be done in the FFAG, with or without an additional buncher ring to obtain very short bunches.

The transverse beam sizes in the near and far term modes would be different. The transverse emittance must be less than ~4mm-mr (rms normalized) to inject into the MI, but would have to be enlarged to ~25mm-mr to avoid space charge limits in the MC mode.

*Scaling solution – 5-plet*

For the neutrino factory international design study (IDS),[6] Rees and Kelliher have developed proton driver designs that include a 3—10 GeV FFAG synchrotron[7]. These solutions could be used for the proton driver, with either the top energy reduced to 8 GeV for Recycler injection or maintained at 10 for MI injection.

The lattice they develop has a series of "pumplet" cells, each of which contains 5 combined-function magnets in an f D F D f configuration (D-magnets are also negative bends) with 4 or 6m long straights. The complete lattice has a circumference of ~400m and an orbit excursion (low to high energy) of only ~10 cm. Lattice parameters are summarized in Table 1.

*Work supported by US DOE under contract DE-AC02-07CH11359
#neuffer@fnal.gov

Table 1: Parameters of a 3-10 GeV FFAG consisting of 16 dual-pumplet cells (dual fDFDf).

| Parameter | Value |
| --- | --- |
| Circumference | 400.8m |
| Tunes ($\nu_x$, $\nu_y$, $\gamma_t$) | 20.8, 10.24, 26.5 |
| f, D, F lengths | 0.5, 0.89, 2.72 m |
| Magnet strengths B, B' (10GeV) | 3.7T, 40T/m |
| Lattice functions ($\beta_{x,m}$, $\beta_{y,m}$, $\eta$) | 27.5, 26, 0.1m |

In adapting this design to the present example, we reduce the peak energy to 8 GeV, while increasing the circumference (increasing number of cells or cell length). This lowers the magnetic field to ~3T or less. More complete dynamic aperture and acceleration studies are needed. Lattice parameters for a single pumplet cell design are shown in Fig. 1.

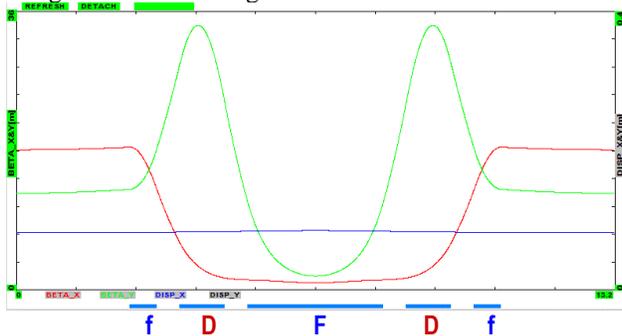

Figure 1: Betatron functions $\beta_x$, $\beta_y$, $\eta$ in a pumplet-type lattice.

## Scaling FFAG - triplet

We have also developed scaling lattices with a triplet geometry. In Table 2 we present parameters for accelerating FFAGs with 948m and 474m circumferences. In these lattices the ~8.6m long cells include 5m straight sections, which should be adequate for injection/extraction kickers plus RF installation. The orbit excursions for these two machines are 33.1 and 47cm for the large-C and small-C, respectively. Betatron functions through a triplet cell are shown in Figure 2.

Dynamic aperture (DA) evaluations of the FFAG have been initiated. Some simulation results are displayed in Fig. 3. They show a very large dynamic aperture for the large-C FFAG; ~3π mm-rad in horizontal and ~10π mm-rad in vertical phase-space. The Small-C FFAG has not yet been evaluated.

Magnetic fields are relatively high in these initial examples, especially for the small-C. The scaling FFAG triplet has relatively equal lengths of positive and negative bending magnets, and the initial design geometry limits the magnet lengths. Future studies will consider variations to reduce these fields.

Table 2: Parameters of 3-8 GeV radial-scaling FFAGs consisting of triplet cells (F D F).

| Parameter | Large-C | Small-C |
| --- | --- | --- |
| Circumference | 948m, 110 cells | 474m, 54 cells |
| Tunes ($\nu_x$, $\nu_y$, $\gamma_t$) | 28.2, 22.4, 19.6 | 14.4, 9.3, 11.6 |
| F, D lengths | 0.78, 1.16m | 0.76, 1.14m |
| Magnet B, B' (8GeV) | 5.5T, 71T/m | 7.4T, 33T/m |
| Lattice functions ($\beta_{x,m}$, $\beta_{y,m}$, $\eta$) | 7, 15, 0.4m | 6.4, 14, 0.56m |
| k value | 383 | 133 |

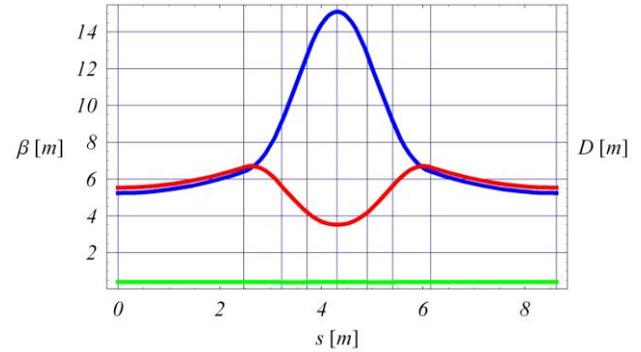

Figure 2: Betatron functions through a cell of the 948m circumference scaling FFAG triplet lattice.

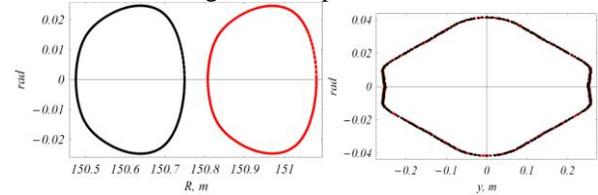

Figure 3: Dynamic apertures for the large-C scaling FFAG. On the left the horizontal dynamic apertures at 3 GeV and 8 GeV are shown, on the right the vertical dynamic apertures are outlined.

## Non-scaling FFAG-triplet

The geometrical constraints are relatively unrestricted if a non-scaling FFAG is used. In particular, this allows a smaller orbit excursion to be obtained, with fewer cells, whilst keeping the magnetic field relatively low. This would have a direct impact on the machine cost, limiting the size of the machine ad aperture in the magnets and RF cavities. A non-scaling FFAG design has been initiated resulting in a preliminary machine parameters shown in Table 3. The non-scaling FFAG has 54 cells, C= 474m and the cell geometry is based on FDF triplet. The magnets have rectangular geometry with parallel faces, which would facilitate its construction and alignment. They would be of the combined function type with multipoles, for chromaticity correction. As in the scaling case, a 5 m long straight section is assumed to be sufficient for injection/extraction requirements. The main difficulty in the non-scaling design is control of the chromaticity, which needs to be close to zero to avoid integer and half integer resonances over the operating range. An initial chromaticity correction using the sextupole and octupole components in the both F and D magnets is shown in Fig. 4. This is yet to be optimised.

Table 3: Parameters of a non-scaling FFAG consisting of FDF type cells.

| Parameter | Value |
|---|---|
| Circumference | 474m, 54 cells |
| Tunes ($\nu_x$, $\nu_y$, $\gamma_t$) | 13.3, 9.3, 50.2 |
| F, D lengths | 0.9, 0.9 m |
| Magnet strengths B, B' (8GeV) | 3.4T, 14T/m |
| Lattice functions ($\beta_{x,m}$, $\beta_{y,m}$, $\eta$) | 7, 17, 0.3m |
| Orbit excursion | ~0.28 m |

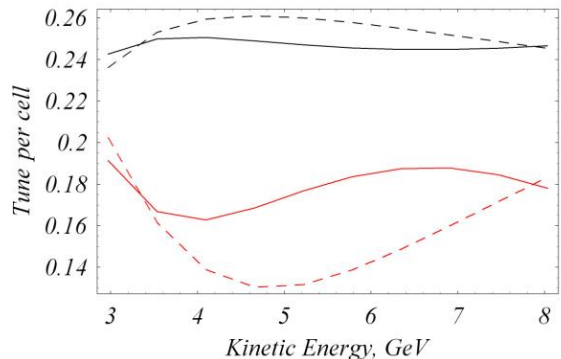

Figure 4: Variation of the horizontal (black) and vertical (red) cell tunes as a function of proton energy in the non-scaling FFAG with chromaticity correction performed using sextupole (dashed lines) or both sextupole and octupole terms (solid lines).

Tracking studies for the non-scaling FFAG have been initiated and suggest that sufficient DA of ~500$\pi$mm-rad can be achieved, but future studies including acceleration and error effects are needed.

## SUMMARY AND FUTURE PLANS

We have presented some initial designs for a 3-8 GeV FFAG, which potentially could be a very interesting solution to deliver high power protons in the multi-MW range, using the Project-X cw linac as an injector. Although the research is not yet mature, the results show that both scaling and non-scaling FFAGs could be implemented. The main issue that needs to be studied in the case of a non-scaling scenario is the tune control by the chromaticity correction. All scenarios require further optimisation, along with tracking studies including errors and the space charge effects. Also the acceleration scenarios need to be developed, together with bunching scenarios that may require a final bunching ring in a MC scenario.